\documentclass[preprint,12pt]{elsarticle}

\usepackage{amssymb}
\usepackage{physics}
\usepackage{amsmath,amssymb}
\usepackage{graphicx}
\usepackage{mathrsfs}
\usepackage{subfigure}

\begin{document}
	
	\begin{frontmatter}

\title{Analytical and numerical calculation of the effect of edge states of the Kane-Mele model on the RKKY interaction }

\author[label1]{Y. Alsayyid}
\author[label1]{J. Ahmadi}
\author[label1]{M. Soltani}
\author[label1]{G. Rashedi}
\author[label2]{Z. Noorinejad}
\address[label1]{Faculty of Physics, University of Isfahan, Isfahan, Iran}
\address[label2]{Department of Physics, Islamic Azad University-Shareza Branch (IAUSH), Shahreza, Iran}

\ead{mo.soltani@sci.ui.ac.ir, rashedi@phys.ui.ac.ir}

\begin{abstract}
In this paper, we investigate the Kane-Mele model and endeavor to demonstrate, through analytical calculations, how the presence of topological edge states influences the RKKY interaction. We illustrate that the effect diminishes as one moves away from the edges. To facilitate our analytical approach, we initially utilize a one-dimensional wire exhibiting linear dispersion for each spin as an approximation to the Kane-Mele model. We examine its impact on the RKKY interaction. Subsequently, we establish a correspondence between the edge states of the Kane-Mele model and a one-dimensional quantum wire model, wherein the coupling strength diminishes with increasing distance from the edges. Finally, we compare the analytical results with numerical findings obtained using the Landauer-Buttiker formulation.
\end{abstract}


\end{frontmatter}

\section{Introduction}
\label{sec.intro}

In 2005, Kane and Mele introduced a simple generalization of the Haldane model of a quantum Hall state without a magnetic field that could conserve time-reversal symmetry \cite{kane2005quantum,kane2005z}. In other words, an electron with an up-spin coupled with an up-direction magnetic field and one with a down-spin coupled with a magnetic field in opposite directions. This type of coupling is called quantum spin-orbit interaction \cite{kargarian2020principles}. Indeed, Kane and Mele provided a model to realize the quantum spin Hall (QSH) insulator with helical edge states, which are protected against weak perturbations by topology \cite{pereg2012inducing,fu2007topological}. It is similar to having two copies of the quantum Hall system, in which the chiral edge states are spin-polarized \cite{ando2013topological}.

The $Z_2$ topological invariant introduced by Kane and Mele was one of the most important results of their model. It led to a new classification of topological insulators, called the $Z_2$ topological insulators or QSH insulators \cite{kane2005z, bernevig2013topological}. They proposed that this time-reversal invariant insulator can be further classified as a trivial insulator with $Z_2 = 0$ or a non-trivial topological insulator with $Z_2 = 1$ \cite{weng2015quantum, hasan2010colloquium}. Although Kane and Mele proposed their model for graphene, Bernevig, Hughes, and Zhang theoretically showed it could be more relevant for two-dimensional $\rm{HgTe/CdTe}$ quantum-well systems \cite{bernevig2006quantum, bernevig2006quan}. It was experimentally confirmed in 2007 by K\"{o}ning \cite{konig2007quantum, konig2008quantum}. After that, these topological classifications of insulators extended to three-dimensional (3D) materials \cite{fu2007topo, moore2007topological}. Angle-resolved photoemission spectroscopy (ARPES) experiments in $\rm{Bi_xSb_{1-x}}$ \cite{hsieh2008topological}, $\rm{Bi_2Se_3}$ \cite{xia2009observation}, and $\rm{Bi_2Te_3}$ \cite{chen2009experimental, hsieh2009tunable} confirmed they are 3D topological insulators.

The Kane-Mele model applies to honeycomb lattices such as graphene, and all Xenes \cite{molle2017buckled} such as silicene \cite{liu2011quantum, liu2011low,ezawa2012valley, pan2014valley}, germanene \cite{liu2011quantum, liu2011low}, and stanine \cite{xu2013large, xu2020new, zhao2020stanene}, etc. Also, jacutingaite ($\rm{Pt_2HgSe_3}$) has been proposed as a novel Kane-Mele QSH insulator \cite{marrazzo2018prediction}. In addition, the transport \cite{ezawa2013quantized, zhou2016tunable} and magnetic \cite{lado2014magnetic, soriano2010spontaneous, wierzbicki2015zigzag} properties of the Kane-Mele model have been investigated.

The model we investigate here is a honeycomb ribbon with zigzag edges composed of the KM model and two magnetic impurities added at the top edge of the ribbon, as shown in Fig.1a. Ruderman-Kittel-Kasuya-Yosida (RKKY) \cite{ruderman1954indirect, kasuya1956electrical, yosida1957magnetic} is an indirect exchange interaction between two magnetic adatoms that is mediated by the helical edge states (Fig.1b) via host itinerant electrons \cite{kittel1968solid}. RKKY interaction could be consist of three terms depending on the spin structure of the host material, Heisenberg-like term, Ising-like term, and Dzyaloshinskii-Moria-like term \cite{imamura2004twisted}. The first term is isotropic collinear and is used for the spin-degenerate system such as graphene \cite{sherafati2011rkky, black2010rkky, parhizgar2013ruderman, szalowski2011rkky}, the second term is anisotropic collinear and is used for spin-polarized systems \cite{sherafati2011rkky, black2010rkky, parhizgar2013ruderman, szalowski2011rkky, valizadeh2016anisotropic}. The third is the anti-symmetric non-collinear term that leads to twisted RKKY interaction used for Rashba spin-orbit coupling \cite{imamura2004twisted, zhu2011electrically, abanin2011ordering} and spin-valley coupling \cite{parhizgar2013indirect, zare2016topological, klinovaja2013rkky} materials. We also have all these three-term interactions for the 3D topological insulator and zigzag silicene nanoribbons \cite{zhu2011electrically, abanin2011ordering}. The single-atomic magnetometry of a pair of magnetic atoms and magneto-transport measurement based on ARPES are several experimental methods for probing RKKY interaction \cite{hindmarch2003direct, zhou2010strength, khajetoorians2012atom}.

The RKKY interactions can be responsible for different magnetic phases, e.g., spiral phases \cite{zare2016topological, christensen2016spiral}, spin-glass \cite{eggenkamp1995calculations, liu1987resistivity}, ferromagnetic \cite{ko2011rkky, ohno1998making, vozmediano2005local, brey2007diluted, priour2004disordered, matsukura1998transport} and antiferromagnetic \cite{minamitani2010effect, hsu2015antiferromagnetic}. Also, it has been proposed to study the topological phase in silicene \cite{zare2016topological, xiao2014ruderman}, edge states of graphene nanoribbon \cite{klinovaja2013rkky}, and 2D topological insulator \cite{zhu2011electrically, shiranzaei2017effect}, decoupled edge modes in phosphorene \cite{islam2018probing}, etc.

One of the most significant properties of topological insulators is the presence of edge states, characterized by a helical edge current in the Kane-Mele topological insulator model. Therefore, examining the RKKY interaction between two magnetic impurities on the edges of a topological insulator nanoribbon should differ substantially from that in the bulk. Consequently, in the topological phase transition, the RKKY interaction for two magnetic impurities on the edge of a nanoribbon undergoes a significant change and can serve as an experimental criterion for the topological phase transition \cite{matsukura2015control}. Additionally, since edge states are robust to impurities, the results obtained for the RKKY interaction are also robust to impurity, and as demonstrated, can be used to construct a one-dimensional Ising model.

We investigate the strength of RKKY interaction between two localized magnetic impurities in the Kane-Mele zigzag ribbon mediated by the helical edge states, first near the edge and then far from the edge. The rest of the paper is organized as follows:

In Section \ref{sec.calculation}, we begin by providing a concise overview of the RKKY approximation. Subsequently, we delve into a model featuring a one-dimensional quantum wire with linear dispersion. This wire interacts with two spin impurities. Through direct calculations, we derive the RKKY approximation's outcome. We evaluate the RKKY interaction by analytically determining the edge states of the Kane-Mele model and employing the findings from Section \ref{sec.calculation}. Additionally, we demonstrate that an increase in distance from the edge leads to a reduction in the RKKY interaction, attributed to the diminishing coupling between edge states and magnetic impurities. Section \ref{sec.Kane-Mele} is dedicated to a comparative analysis with numerical computations, utilizing the Landauer-Buttiker formulation to acquire Green’s function. A summary and conclusion are presented in Section \ref{sec.conclu}.

\section{Calculation of RKKY interaction in the presence of a topological current-carrying quantum wire}
\label{sec.calculation}
In this section, our objective is to calculate the RKKY approximation for the simplified model depicted in Fig.\ref{fig.2}. In this model, we make the assumption that two spin impurities are located at a distance $d$ from each other within a quantum wire that carries a current with a specific helicity. Utilizing the Green functions of this wire, we proceed to compute the RKKY interaction. Moving forward to the next section, we will demonstrate the resemblance between the edge mode of the Kane-Mele model and a quantum wire with directional spin.

\subsection{The RKKY interaction }
\label{sec.The RKKY} 
If two spin impurities are located within the Fermi Sea, an indirect magnetic interaction arises between them due to their interaction with the shared electron environment. This phenomenon, first introduced by Ruderman-Kittel-Kasuya-Yosida (RKKY), is known as RKKY interaction. In this model, it is demonstrated that when two spin impurities are positioned within the same environment, each impurity interacts with the electrons in the following manner
\begin{equation}\
 \label{eq.1}
   H_{int}=J_c\sum_{i=1,2}\mathbf{\hat{S}}_e(r_i)\cdot\mathbf{\hat{S}}_i,           
 \end{equation}
where,$J_c$ denotes the coupling constant between conduction electrons and $\mathbf{S}_i$ represents the spin operator of the $i$th impurity. For one dimentional model, it can be expressed as follows
\begin{equation}\
 \begin{aligned}
  \mathbf{\hat{S}_1}&=\frac{\hbar}{2}\hat{\sigma}_1 \delta(x)\\
  \mathbf{\hat{S}_2}&=\frac{\hbar}{2}\hat{\sigma}_2 \delta(x-L),
 \end{aligned}
 \end{equation}
$\sigma$ signifies the Pauli spin matrix. In Eq.(\ref{eq.1}), $\mathbf{S}_e(r)=\frac{\hbar}{2}\sum_{j}\delta(r_j-r)\sigma_j$ is the spin operator with $r_j$ and $\sigma_j$ denoting the position and the spin operators vector of the $j$th electron, respectively. Through second-order perturbation theory, it can be shown that this interaction leads to an indirect interaction between two impurities, which can be written as follows:
\begin{equation}\
 \label{eq.2}
   H_{RKKY}=C\mathbf{\hat{S}}_1\cdot\boldsymbol{\chi}\cdot\mathbf{\hat{S}}_2.           
 \end{equation}
Where,$\hat{S}_1$ and $\hat{S}_2$ are located at positions $r$ and $r^\prime$, respectively. Here, $K$ and $L$ denote the sublattice index on which magnetic impurities are placed, and $J=C \boldsymbol{\chi}^{KL}$ is the strength of the exchange coupling between two impurities where $C=(\frac{J_C\hbar}{2})^2$. The spin susceptibility of the system, denoted by ${\chi^{KL}}$ can be calculated as follows

 \begin{equation}\
 \label{eq.3}
   \mathbf{\chi}^{KL}=\frac{-2}{\pi}\Im \int_{-\infty}^{\varepsilon_f}d\varepsilon \Tr[\hat{\sigma}^K \hat{G}(r_1,r_2,\varepsilon) \hat{\sigma}^L \hat{G}(r_2,r_1,\varepsilon)].         
 \end{equation}

In this relation, $\hat{\sigma}^L$ represents the Pauli spin matrices, and $\hat{G}(r_1,r_2,\varepsilon)$ is a (2 × 2) matrix of the single-particle retarded Green’s functions in the spin space. The trace is taken over the spin degree of freedom. Here, $\varepsilon_f$ denotes the Fermi energy, which we assume to be zero. As previously mentioned, we begin our energy measurement from the upper edge of the conduction band in the Kane-Mele model.

\subsection{Calculation of RKKY interaction in the presence of helical quantum wire  }
\label{sec.helical quantum}
As shown in Eq. (\ref{eq.3}), to calculate the RKKY interaction, we first need to compute the Green's function of an electron. For the one-dimensional system in the above model, the Hamiltonian of the quantum wire is given by
\begin{equation}\
 \label{eq.4}
   H_{wire}^{(k)}=\int_{-\alpha}^{\alpha}dk \hbar v^\prime_f \hat{\sigma}_z k\ket{k}\bra{k}.          
 \end{equation}

Where $v^\prime_f=\frac{v_f}{a}$, in which $v_F$ and $a$ are the Fermi velocity and lattice constant, respectively. Also, we assume that $k$ is the dimensionless quantity where $-\alpha<k<\alpha$. We will obtain the values of $a$ and $\alpha$ in the next section.

As llustrated in Fig.\ref{fig.2}, electrons with up and down spins move to the right and left, respectively. In the upcoming section, we will demonstrate that the edge states in the Kane-Mele model exhibit similar behavior.
 Based on Hamiltonian (\ref{eq.4}), the Green's function can be defined as follows
 \begin{equation}\
 \label{eq.6}
 \begin{aligned}
   G_{\uparrow\uparrow}(x,x^\prime,\varepsilon)&=\int\frac{\bra{x}\ket{k}\bra{k}\ket{x^\prime}}{\varepsilon-\varepsilon_k+i0^+}dk\\
                         &= \int\frac{e^{ik(x-x^\prime)}}{\varepsilon-\varepsilon_k+i0^+}dk=\int\frac{e^{ik(x-x^\prime)}}{\varepsilon-\hbar v_fk+i0^+}dk.      
 \end{aligned}
 \end{equation}
If $x > x^\prime$, to evaluate the integral (\ref{eq.6}) using the contour integration method, we close the integration contour in the lower half of the complex plane. Given that the first-order pole is also in the lower half complex plane, its value will be $\frac{\varepsilon}{\hbar v^\prime_f}$. On the other hand, if $x < x^\prime$, we close the integration contour in the upper half of the complex plane. In this case, an exterior pole exists outside the contour, resulting in a vanishing integral. In other words
\begin{equation}\
\label{eq.7}
\begin{aligned}
G_{\uparrow\uparrow}(x,x^\prime,\varepsilon)&=\begin{cases}
     \frac{2\pi i}{\hbar v^\prime_f} e^{i\frac{\varepsilon}{\hbar v^\prime_f} (x-x^\prime)} & x>x^\prime \\ 0 & x< x^\prime,         
   \end{cases}\\
   \end{aligned}
\end{equation}
similarly
\begin{equation}\
\label{eq.8}
\begin{aligned}
G_{\downarrow\downarrow}(x,x^\prime,\varepsilon)&=\begin{cases}
     0 & x>x^\prime \\ \frac{2\pi i}{\hbar v^\prime_f}e^{-i\frac{\varepsilon}{\hbar v^\prime_f}(x^\prime-x)} & x< x^\prime.         
   \end{cases}\\
   \end{aligned}
\end{equation}

We must note that in these calculations, the range of the allowable wave number $k_n$ that we give to the electrons of the quantum wire must be in the range $\alpha$, the value of $\alpha$ will be calculated in the next section. The reason for this limitation is that the edge states of the Kane-Mele model are defined in only one allowable range in terms of $k$ space. Now we can simply calculate the relation (\ref{eq.3}) using the obtained Green’s functions.
Since $G_{\downarrow\uparrow} = G_{\uparrow\downarrow} = 0$, the Green’s function will take the form of
\begin{equation}\
\begin{aligned}
G_{(x-x^\prime)}&=\begin{cases}
     \left(
                        \begin{array}{cc}
                          0 & 0 \\
                          0 & G_{\uparrow\uparrow} \\
                        \end{array}
                      \right) & x>x^\prime \\ \left(
                       \begin{array}{cc}
                         G_{\downarrow\downarrow} & 0 \\
                         0 & 0 \\
                       \end{array}
                     \right) & x< x^\prime.
   \end{cases}\\
\hat{G}(x-x^\prime)&= \left(
                        \begin{array}{cc}
                          0 & 0 \\
                          0 & G_{\uparrow\uparrow} \\
                        \end{array}
                      \right)
for  x>x^\prime\\
\hat{G}(x-x^\prime)&=\left(
                       \begin{array}{cc}
                         G_{\downarrow\downarrow} & 0 \\
                         0 & 0 \\
                       \end{array}
                     \right)
for  x<x^\prime
\end{aligned}
\end{equation}
According to Eq.(\ref{eq.3}) and the form of Green’s functions, we conclude that the trace will be non-zero only when $K=L=z$. Therefore, we expect the RKKY interaction to be as follows
\begin{equation}\
 \label{eq.9}
   H_{RKKY}=J^\prime \hat{S}_{1Z}\hat{S}_{2Z},          
\end{equation}

where $J^\prime=C \boldsymbol\chi^{z,z}$, indicating that the interaction takes on the form of the Ising interaction.
\begin{equation}\
 \label{eq.10}
 \begin{aligned}
   \boldsymbol\chi^{z,z}&=-\frac{2}{\pi}\Im\int_{-\alpha \hbar v^\prime_f}^{0}d\varepsilon\Tr[\hat{\sigma}^z \hat{G}_{\uparrow\uparrow}(x,x^\prime,\varepsilon)\hat{\sigma}^z \hat{G}_{\downarrow\downarrow}(x,x^\prime,\varepsilon)]\\
   \boldsymbol\chi^{z,z}&=-\frac{2}{\pi}\Im\int_{-\alpha}^{0}d\varepsilon^\prime \frac{4\pi^2}{\hbar v^\prime_f}e^{2i \varepsilon^\prime(x-x^\prime)}
   \end{aligned}         
\end{equation}

where $\varepsilon^\prime= \frac{\varepsilon}{\hbar v^\prime_f}$, we arrive at the following straightforward conclusion
\begin{equation}\
 \label{eq.11}
 \begin{aligned}
   \boldsymbol\chi^{z,z}&=-\frac{2}{\pi}\int_{-\alpha}^{0}d\varepsilon^ \prime   \frac{4\pi^2}{\hbar v^\prime_f} \sin(2 \varepsilon^\prime(x-x^\prime))\\
   &=(\frac{4\pi}{\hbar v^\prime_f(x^\prime-x)})(\cos( 2\alpha(x^\prime-x))-1)\\
   &=(\frac{4\pi}{\hbar v^\prime_f(x^\prime-x)})\sin^2( 2\alpha(x^\prime-x))
   \end{aligned}         
\end{equation}

\section{Calculation of edge states of Kane-Mele model and calculation of interaction strength between spin impurities and edge states }
\label{sec.Kane-Mele}
In this section, we demonstrate that, through a well-founded approximation, we can effectively emulate the Hamiltonian of the hypothetical quantum wire introduced in the previous section using the edge state of the Kane-Mele model.Then, by calculating the wave function of the edge states, we calculate the interaction strength between the edge state and the spin impurity and approximate $v^\prime_f$ and $\alpha$ for the relation (\ref{eq.4}). Finally, we compare the analytical results with the numerical results. To calculate the Green's function as well as the $\alpha$ and $v_f$ coefficients, one can utilize the results from the reference \cite{rahmati2023explicit}. It presents an analytical method to obtain highly accurate dispersion and edge-state wavefunction results.

As established in articles \cite{rahmati2023explicit} and \cite{sadeghizadeh2023rigorous}, the exploration of edge states within the Haldane model necessitates a departure from the conventional approach seen in the graphene model. Unlike the direct mapping characteristic of the graphene model, associating each wave vector $ k_x $ with an SSH chain, the Haldane model requires a conceptualization resembling an SSH ladder. In this conceptualization, each zigzag corresponds to a dual-atom mapping. Subsequently, the focus shifts to identifying specific localized states within this ladder model. An intriguing observation unfolds—despite the ladder mapping, only one distinct localized edge state prevails in the Haldane model. As outlined in \cite{sadeghizadeh2023rigorous}, these localized edge states exclusively manifest on the odd zigzags. The rigorous proof for the aforementioned claims is meticulously detailed in \cite{sadeghizadeh2023rigorous}, allowing interested readers to delve into the computational intricacies. Our discussion here is confined to presenting the outcomes.

\begin{figure}
    \centering
    \subfigure[]{\includegraphics[width=0.6\textwidth]{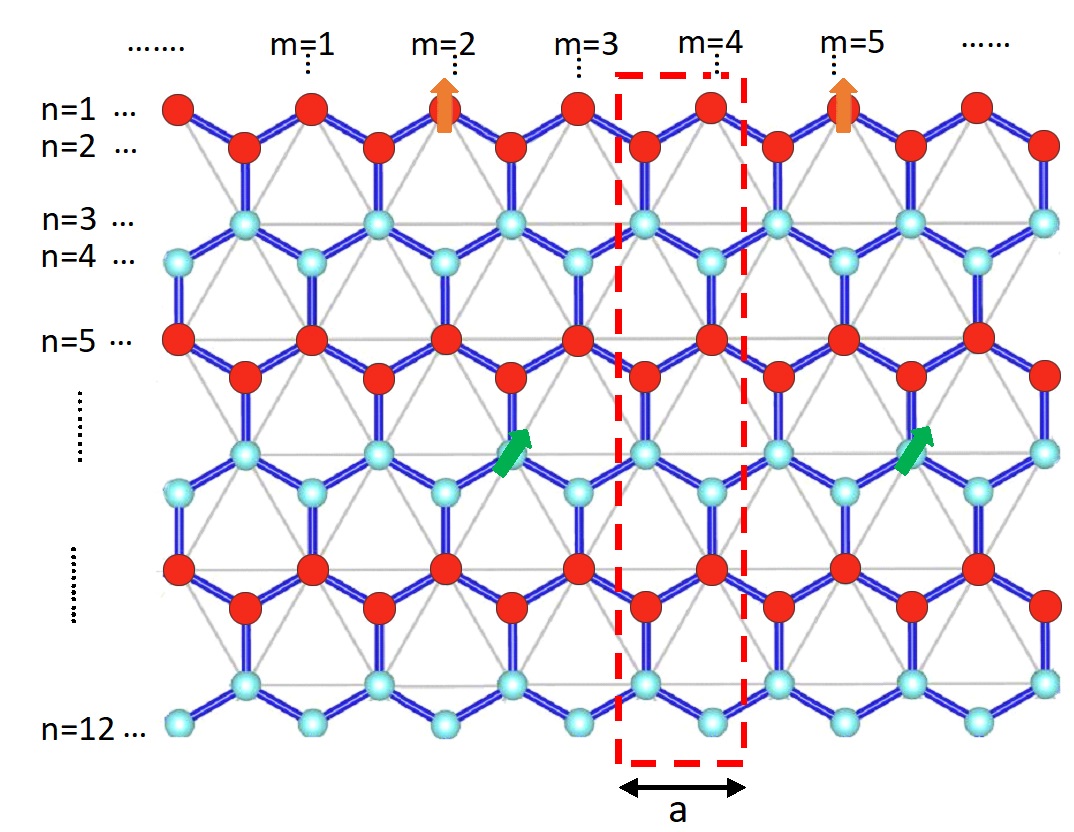}}
    \subfigure[]{\includegraphics[width=0.6\textwidth]{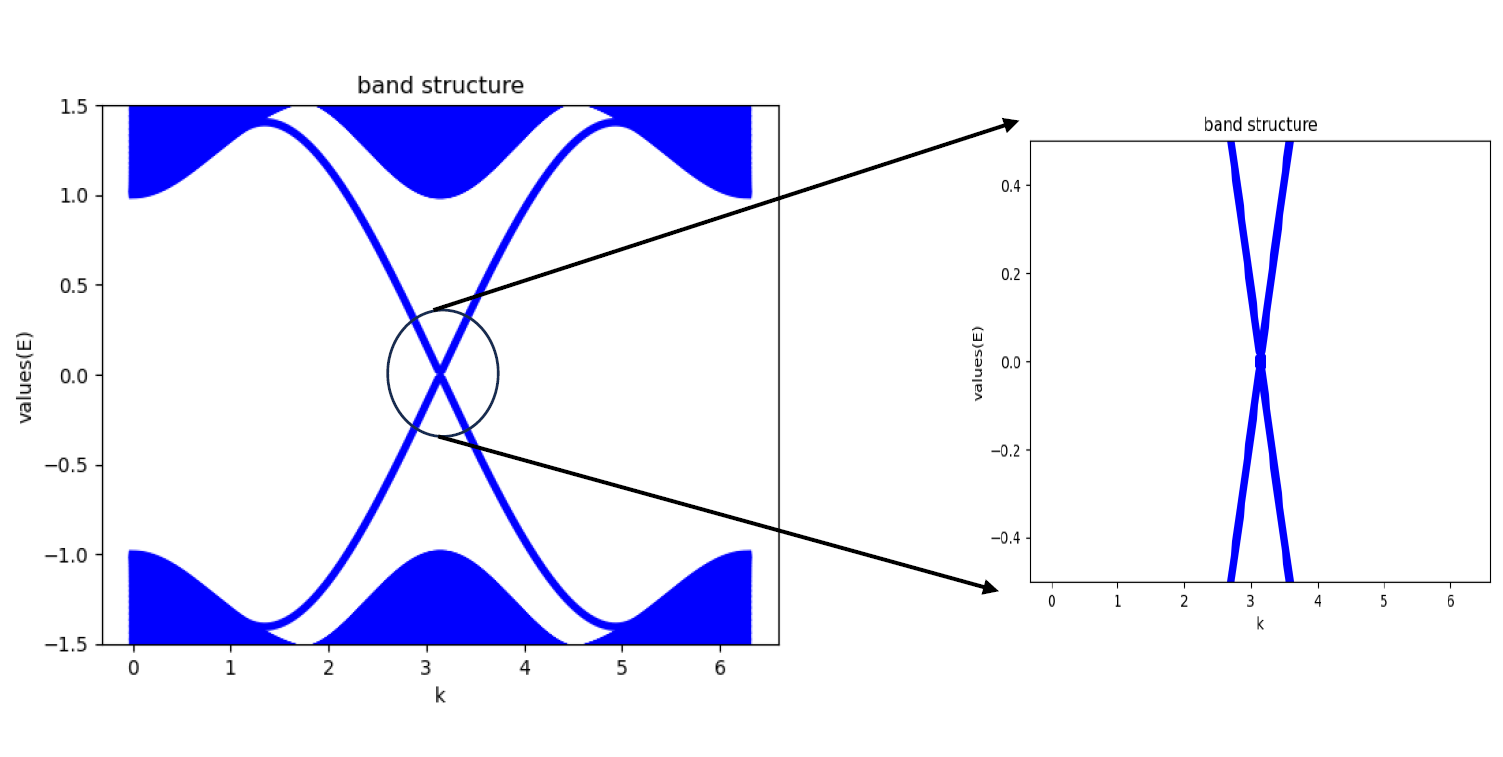}}
    \caption{(a) illustrates a nanoribbon with zigzag edges in the Kane-Mele model, where two spin impurities are located at sites $n_1$ and $n_2$. The wavefunction of edge states is nonzero on the blue zigzags, while it is zero on the red zigzags. (b)Illustrates the nanoribbon band structure of the Kane-Mele model with $t=1$ and $\lambda_{so}=0.3$. } 
    \label{fig.1}
\end{figure}

\begin{figure}[h]
  \centering
  \includegraphics[width=7cm]{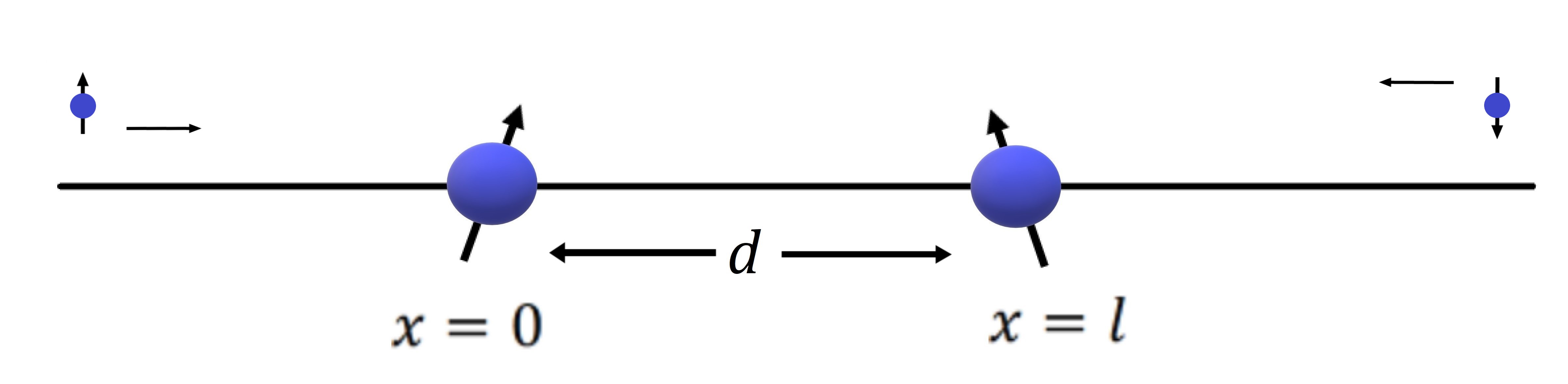}
  \caption{Schematic of a quantum wire carrying spin currents with two magnetic impurities localized in the $x=0$ and $x=l$.}
  \label{fig.2}
\end{figure}

1-	Wavefunction: As shown in Fig.\ref{fig.1}a, the wavefunction takes zero and non-zero values on the  red and blue zigzags, respectively. In other words, the wavefunction is zero on even zigzags and non-zero on odd zigzags. The edge wavefunction on odd zigzag for atoms of type $A$ and $B$ is expressed as follows,
\begin{equation}\
 \label{eq.13}
\begin{aligned}
   \psi_a^n(k_x)&=\gamma(k_x) \Lambda^n(k_x) \sin(\frac{\theta_{k_x}}{2}),\\
   \psi_b^n(k_x)&=\gamma(k_x) \Lambda^n(k_x) \cos(\frac{\theta_{k_x}}{2}),          
\end{aligned}
\end{equation}

where 

\begin{equation}\
\begin{aligned}
    \Lambda^n(k_x) &=1+\frac{1}{2}(\frac{t}{2\lambda_{so}\sin(\frac{k_x}{2})})^2-\\& \frac{t}{2\lambda_{so}\sin(\frac{k_x}{2})}  \sqrt{\frac{1}{4}(\frac{t}{2\lambda_{so}\sin(\frac{k_x}{2})})^2+1},\\
&\theta_{k_x}=\tan^{-1}(\frac{4\lambda_{so}\sin(\frac{k_x}{2})}{t}),\\
&\gamma(k_x)=\sqrt{1-\Lambda^2(k_x)}.
\end{aligned}
\end{equation}
2-	Edge-state dispersion: Zigzag edge states possess the following dispersion. 

\begin{equation}\
 \label{eq.15}
   \epsilon^{\pm}=\mp \frac{3t\cos(\frac{k_x}{2})}{\sqrt{\frac{t^2}{16\lambda_{so}^2\sin^2(\frac{k_x}{2})}+1}} ,          
\end{equation}

Moreover, these edge states with up and down spins exhibit distinct dispersion behaviors, featuring positive and negative group velocities, respectively. Notably, both up and down spin states possess zero energies at $k_x=\pi$. Also, Eq(\ref{eq.15}) is valid in the range of $\pi-\alpha<k_x<\pi+\alpha$, we focus on $\alpha$ values within the interval where the edge states and bulk states intersect. To calculate $\alpha$, we utilize the known energy gap of the Kane-Mele model as follows

\begin{equation}\
 \label{eq.16}
   \epsilon_g=6\sqrt{3} \lambda_{so} ,          
\end{equation}

Thus, $\alpha$ can be written as follow

\begin{equation}\
 \label{eq.17}
   \frac{3t\cos(\frac{\pi}{2}+\frac{\alpha}{2})}{\sqrt{\frac{t^2}{16\lambda_{so}^2\sin^2(\frac{\pi}{2}+\frac{\alpha}{2})}+1}}= 3\sqrt{3} \lambda_{so},          
\end{equation}

To precisely determine $\alpha$, numerical solutions are essential for the given equation. However, rather than directly solving this equation, we shift our focus to the group velocity $ \frac{\partial \epsilon}{\partial k} \big|_{k=\pi}=v_f \approx \frac{9\sqrt{3}\lambda_{so}}{\pi} $. Notably, in the proximity of $ k=\pi $, the dispersion exhibits a robust linear approximation. Consequently, in our $\alpha$ calculations, we adopt the assumption $ v_f\alpha=\frac{\epsilon_g}{2} $, leading to $ \alpha=\frac{\pi}{3} $. This assumption aligns seamlessly with our numerical results \cite{rahmati2023explicit}, ensuring a robust correspondence.

In this paper, we aim to investigate the RKKY interaction, as depicted in Fig.\ref{fig.2}. As shown in this figure, two spins are placed on the zigzag edges of a nanoribbon in the Kane-Mele model. Therefore, to solve Eq.(\ref{eq.3}), we require the following Green's function

\begin{equation}\
 \label{eq.18}
   \hat{G}=\int_{\pi-\alpha}^{\pi+\alpha}{\frac{\ket{k}\bra{k}}{\epsilon-\epsilon_k+i0^+}dk},          
\end{equation}

By utilizing the definition of the Green's function and Eq.(\ref{eq.13}), with some calculations, we obtain at the following results

\begin{equation}\
\label{eq.19}
\begin{aligned}
G^{\uparrow\uparrow}_{n_1, n_2}(\varepsilon)&=\begin{cases}
     \frac{2\pi i}{\hbar v_f} e^{i k_\varepsilon (n_1 - n_2)}\psi^2_1 (k_\varepsilon) & n_1>n_2 \\ 0 & n_1<n_2 ,

   \end{cases}\\
G^{\downarrow\downarrow}_{n_1, n_2}(\varepsilon)&=\begin{cases}
     0 & n_1>n_2\\ \frac{2\pi i}{\hbar v_f} e^{i k_\varepsilon (n_1 -n_2)}\psi^2_1 (k_\varepsilon) & n_1< n_2,         
   \end{cases}
   \end{aligned}
\end{equation}

where the mentioned $k_\varepsilon$ satisfies the following equation

\begin{equation}\
 \label{eq.20}
   \epsilon= \frac{3t\cos(\frac{k_\varepsilon}{2})}{\sqrt{\frac{t^2}{16\lambda_{so}^2\sin^2(\frac{k_\varepsilon}{2})}+1}} ,    
\end{equation}

\begin{equation}\
 \label{eq.21}
\hbar v_f=   \frac{\partial \epsilon}{\partial k} |_{k_\varepsilon}\approx \frac{\partial \epsilon}{\partial k} |_{k=\pi}= \frac{9\sqrt{3}\lambda_{so}}{\pi}  
\end{equation}
In Eq.(\ref{eq.21}), we disregard changes in group velocity, leveraging the assumption of linear dispersion. As seen in Eq.(\ref{eq.19}), the obtained Green's function closely resembles the Green's function derived in Eqs.(\ref{eq.7}) and (\ref{eq.8}). The only main difference is the appearance of a coefficient $\psi^2_1(k_\varepsilon)$ in Eq.(\ref{eq.19}). Other calculations closely align with the results presented in Eq.(12); in other words, we have

\begin{equation}\
   \chi^{z,z}=\frac{-2}{\pi} \Im \int_{-\alpha }^{0}{d\varepsilon (\frac{2\pi}{\hbar v_f})e^{\frac{2i\pi}{\hbar v_f}(n-n^{\prime})} \psi^2(k_{\varepsilon})},          
\end{equation}

If we ignore variations in $\psi^2_1(k_{\varepsilon})=\psi^2(k_{\varepsilon})$  where  $\varepsilon=0$, we arrive at the following result for the magnetic interaction

\begin{equation}\
 \label{eq.22}
   \chi^{z,z}=\frac{2\pi \psi^2}{\hbar v_f(n-n^{\prime})} \sin^2( 2\alpha  (n-n^{\prime}) ,          
\end{equation}

\begin{figure}
    \centering
    \subfigure[]{\includegraphics[width=0.45\textwidth]{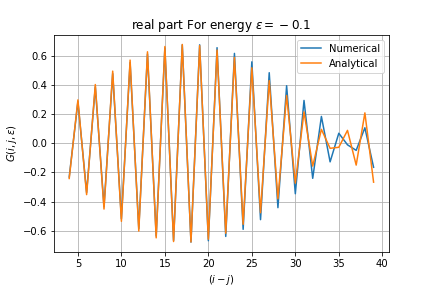}}
    \subfigure[]{\includegraphics[width=0.45\textwidth]{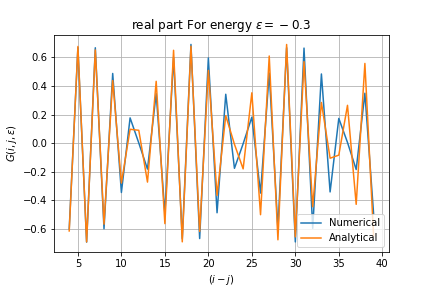}}
    \subfigure[]{\includegraphics[width=0.45\textwidth]{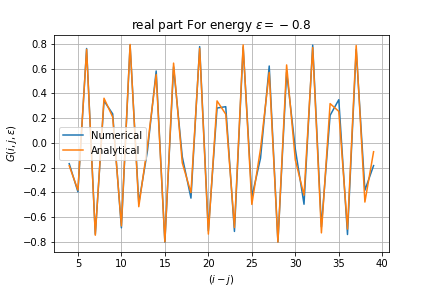}}
    \caption{Comparison of numerical and analytical results of the imaginary part of the Green's function Eq.(\ref{eq.25}) for cases (a) $\varepsilon=-0.1t$, (b) $\varepsilon=-0.3t$, and (c) $\varepsilon=-0.8t$}.
    \label{fig.3}
\end{figure}

As we employed various approximations to calculate the Green's function, we compare the numerical and analytical results for the Green's function in Eq.(\ref{eq.19}), as illustrated in Fig.\ref{fig.3}. By numerical calculations, we mean that we can consider the Green's function as follows 
\begin{equation}\
       G^{\uparrow\uparrow}_{n_1, n_2}(\varepsilon)=i g_n(\varepsilon)e^{ik_\varepsilon(n_1-n_2)} .                  
\end{equation}

\begin{figure}
    \centering
    \subfigure[]{\includegraphics[width=0.6\textwidth]{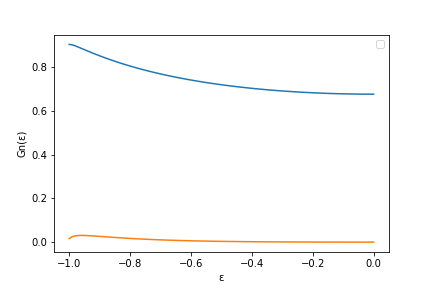}}
    \subfigure[]{\includegraphics[width=0.6\textwidth]{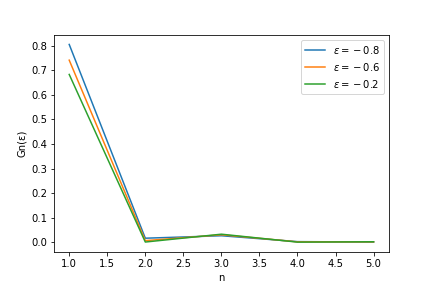}}
    \caption{(a) $g_{00}$ is shown for different $\varepsilon$ values on the first zigzag edges with a continuous line plot and on the second zigzag edges with a scatter plot. The coefficients are similar to those in Fig.\ref{fig.1}a (b)Numerical results of the Green function for various energies as a function of distance from edge states.} 
    \label{fig.4}
\end{figure}

Consequently, the real and imaginary parts of Eq(\ref{eq.21}) will be as follows
\begin{equation}\
      \Re G^{\uparrow\uparrow}_{n_1, n_2}(\varepsilon)= g_0(\varepsilon)\sin{k_\varepsilon(n_1-n_2)} ,                  
\end{equation}
and
\begin{equation}\
\label{eq.25}
      \Im G^{\uparrow\uparrow}_{n_1, n_2}(\varepsilon)= g_0(\varepsilon)\cos{k_\varepsilon(n_1-n_2)} .                  
\end{equation}
Where $g_n(\varepsilon)=g_{00}^2(\varepsilon)$.
The Fig.\ref{fig.4} reveal an excellent agreement between the analytical and numerical outcomes, validating the accuracy of the analytical approach proposed in this article. Additionally, the influence of $\psi_1^2(k_0)$ on the Green's function is examined in Fig.\ref{fig.3} and \ref{fig.4}a. As evident, the Green's function remains non-zero and nearly constant on the first zigzag (indicating a good second approximation in Eq.(\ref{eq.13}), while it suddenly tends to zero on the second zigzag, supporting the analytical results from reference \cite{rahmati2023explicit}.

Another crucial observation is that the RKKY interaction resulting from the edge-state interaction in the Kane-Mele model is significant only in the vicinity of the edges. As the edge-state wavefunctions rapidly approach zero away from the edges (i.e., $\alpha_n(k)$ in Eq.(\ref{eq.13}) becomes small), the RKKY interaction becomes negligible beyond the proximity of the edges. Given the numerical values employed in this study, the $\alpha$ parameter is on the order of 0.15. Consequently, the interaction strength within the third zigzag is a mere 0.03 relative to the interaction strength at the edge. This numerical characterization of the phenomenon is visually depicted in Fig.\ref{fig.4}b. Considering Eq.(\ref{eq.22}), we conclude that the RKKY interaction at the edges of the Kane-Mele model exhibits an Ising-like behavior, decaying as $\frac{1}{r}$. This distinctive interaction may hold promise for the implementation of Ising spin chains.

\section{Conclusion }
\label{sec.conclu}

In this article, we initially investigated the RKKY interaction between two magnetic impurities in the presence of a quantum wire carrying a current with helical edge states . We demonstrated that in such a system, the RKKY interaction follows an Ising-like behavior, decaying as $\frac{1}{r}$. Furthermore, we showed that in the presence of edge states in the Kane-Mele model, the results closely resemble those of the aforementioned model. However, the main difference is that as one moves away from the edges, the RKKY interaction rapidly tends toward zero. The proposed model could have practical applications in constructing Ising spin chains, making it suitable for various research and technological developments.

\bibliographystyle{elsarticle-num}
\bibliography{reference1}

\end{document}